\documentclass[twocolumn]{aastex62}

\usepackage{amsmath}
\usepackage{textcomp}
\usepackage{soul}

\def \h2{{\rm H_{2}}}

\def \dn4000{D_{{\rm n}}(4000) }

\def \wone{\textit{W1}}
\def \WISE{\textit{WISE}}
\def \NEOWISE{\textit{NEOWISE}}
\def \wtwo{\textit{W2}}

\accepted{for publication in ApJL, August 2019}

\shorttitle{How to Find Variable AGN with Machine Learning}
\shortauthors{Faisst, Prakash, Capak, \& Lee}
\begin{document}

\title{\sc \large How to Find Variable Active Galactic Nuclei with Machine Learning}

\correspondingauthor{A. Faisst / A. Prakash
}
\email{afaisst@ipac.caltech.edu / aprakash@ipac.caltech.edu
}

\author[0000-0002-9382-9832]{Andreas L. Faisst}
\affil{IPAC, California Institute of Technology
1200 E California Boulevard, Pasadena, CA 91125, USA}

\author[0000-0003-4451-4444]{Abhishek Prakash}
\affil{IPAC, California Institute of Technology
1200 E California Boulevard, Pasadena, CA 91125, USA}

\author[0000-0003-3578-6843]{Peter L. Capak}
\affil{IPAC, California Institute of Technology
1200 E California Boulevard, Pasadena, CA 91125, USA}
\affil{Cosmic Dawn Center (DAWN), Copenhagen, Denmark}

\author[0000-0003-1954-5046]{Bomee Lee}
\affil{IPAC, California Institute of Technology
1200 E California Boulevard, Pasadena, CA 91125, USA}

\begin{abstract}
Machine-learning (ML) algorithms will play a crucial role in studying the large datasets delivered by new facilities over the next decade and beyond. Here, we investigate the capabilities and limits of such methods in finding galaxies with brightness-variable active galactic nuclei (AGN). Specifically, we focus on an unsupervised method based on self-organizing maps (SOM) that we apply to a set of nonparametric variability estimators. This technique allows us to maintain domain knowledge and systematics control while using all the advantages of ML.
Using simulated light curves that match the noise properties of observations, we verify the potential of this algorithm in identifying variable light curves.
We then apply our method to a sample of $\sim8300$ \textit{WISE} color-selected AGN candidates in Stripe 82, in which we have identified variable light curves by visual inspection. We find that with ML we can identify these variable classified AGN with a purity of $86\%$ and a completeness of $66\%$, a performance that is comparable to that of more commonly used supervised deep-learning neural networks. 
The advantage of the SOM framework is that it enables not only a robust identification of variable light curves in a given dataset, but it is also a tool to investigate correlations between physical parameters in multi-dimensional space $-$ such as the link between AGN variability and the properties of their host galaxies. Finally, we note that our method can be applied to any time-sampled light curve (e.g., supernovae, exoplanets, pulsars, and other transient events).

\end{abstract}
\section{Introduction} \label{sec:intro}

Over the next decade, new facilities will deliver a tremendous amount of data to study astrophysical phenomena. In order to trawl through these large data volumes, fast, automated, and efficient methods are needed. Machine-learning (ML) algorithms are a powerful tool to identify, classify, characterize, and visualize astronomical objects and the correlations of their physical properties in multi-dimensional parameter space. They are already being used to derive photometric redshifts and other physical properties of galaxies \citep[][]{MASTERS15,KRAKOWSKI16,SPEAGLE17a,SPEAGLE17b,SIUDEK18,BONJEAN19,DAVIDZON19,HEMMATI19,MASTERS19,TURNER19}, as well as to classify light curves of supernovae and to identify other galactic transient events \citep{LOCHER16,CHARNOCK17,SESAR17,CARRASCO18,HINNERS18,SOOKNUNAN18,AQUIRRE19,MUTHUKRISHNA19a,MUTHUKRISHNA19b}.

Here, we investigate the capabilities of ML algorithms in finding galaxies with variable active galactic nuclei (AGN).
Powered by the accretion of matter onto supermassive black holes (SMBH) residing in the center of galaxies, AGN shape the evolution and structure of their host galaxy through various feedback mechanisms \citep{BOWER06,CATTANEO06,CROTON06,SIJACKI07,hopkins_agn_2012,DUBOIS13}.
Measuring the number-density and rate of occurance of AGN therefore enables the study the formation and growth of SMBHs and their host galaxies across cosmic time \citep{peterson_1997}.
Due to different ``feeding mechanisms,'' AGN exhibit variations in brightness over a range of wavelengths on timescales ranging from minutes to years \citep{fitch_1967}. Variations on short timescales are likely caused by disk instabilities \citep{kawaguchi_1998}, while variations on longer timescales are dominated by the fueling of gas into the nuclear regions and regulation through feedback processes \citep[e.g.,][]{hopkins_agn_2012}.
The study of AGN variability levels therefore adds another dimension and allows us to learn about the internal processes such as inflows and outflows and the size of accretion disks around SMBHs \citep{shields_1978}.

In this Letter, we demonstrate the capabilities of self-organizing maps \citep[SOM,][]{KOHONEN82,KOHONEN90}, an un-supervised ML algorithm, in identifying AGN displaying long-term brightness variability. This algorithm has been widely used in the past, for example to study radio galaxies \citep{TORNIAINEN08,RALPH19}, variable stars \citep{BRETT04, ARMSTRONG16}, and exoplanet transit curves \citep{ARMSTRONG17} as well as to derive photometric redshifts \citep{CARRASCO14,MASTERS15,HEMMATI19,HEMMATI19b} and to classify gravitational waves \citep{RAMPONE13}.
We apply the SOM algorithm to a set of nonparametric variability estimators, which allows us to maintain domain knowledge of the data properties encapsulated in these estimators (e.g., noise, sampling rate, and selection function), and offers the flexibility of learning-based non-linear classifications that can optimally combine these estimators for classification. Such techniques will be especially valuable for extrapolating knowledge from the deep and well sampled parts of future surveys, such as LSST, Euclid, and \textit{WFIRST}, to the wide and shallow parts with poor sampling.
We emphasize that our work serves as a proof of concept, and the methods described here can be further refined and extended (e.g., to flux variability at multiple wavelengths).

The \textit{WISE} AGN sample, light curves, and different variability estimators are presented in Section~\ref{sec:data}. In Section~\ref{sec:classification}, we apply the SOM algorithm to our sample and compare its performance to a deep-learning regression fitting method. We conclude in Section~\ref{sec:end}.
Magnitudes are expressed in the AB system \citep{oke_gunn_1983} unless stated otherwise. We use a standard $\Lambda$CDM cosmology with $H_{0}$=70 km s$^{-1}$ Mpc$^{-1}$, $\Omega_{\rm M} = 0.3$, and $\Omega_\Lambda = 0.7$.

\section{\textit{WISE} AGN Sample} \label{sec:data}

\subsection{Sample Selection and Light Curves}\label{sec:wise_agn}

The AGN used in this study are color selected from a sample of $14,000$ AGN in the 270\,deg$^{2}$ Sloan Digital Sky Survey (SDSS) Stripe 82 field \citep{JIANG14}. The details are outlined in \citet{PRAKASH19}, a summary is provided in the following.

The color selection follows the criteria of \citet{stern_agn_2012}, using the $\wone$~and $\wtwo$ filters of the Wide-field Infrared Survey Explorer \citep[WISE,][]{Wright2010} centered on $3.4\,{\rm \mu m}$ and $4.6\,{\rm \mu m}$. Specifically, we apply a cut at $\wone-\wtwo>0.8\,{\rm mag_{Vega}}$ and restrict ourselves to $\wtwo>15\,{\rm mag_{Vega}}$. About $50\%$ of the candidates have confirmed redshifts and lie between $0.1 < z < 2.2$ with a median of $z\sim0.5$.

Once the AGN candidates are identified, their time-sampled photometry is measured on \textit{WISE} $\wone$ single exposure (Level 1b) images in apertures of $6\arcsec$. Bad pixels indicated by the bad pixel mask are excluded. Several standard stars are used to correct the photometry for aperture losses. The light curves are generated using only high-quality frames well separated from the South Atlantic Anomaly and bright moon light by selecting $\texttt{qual\_frame}=10$, $\texttt{SAA\_SEP}>0$, and $\texttt{MOON\_SEP}>24$ as suggested by the \WISE~team\footnote{\url{http://wise2.ipac.caltech.edu/docs/release/allsky/expsup/sec2_4b.html}}.

The final light curves include all data from \WISE~ and \NEOWISE~\citep{neowise} over the past 10 years. Note that a $\sim3.5$ year gap around MJD 55725 arises during the hibernation period of the telescope.
\textit{WISE} observed a single patch of sky multiple times during each visit, leading to multiple observations within typically $\sim 1-2\, {\rm days}$. Since our focus here is on long-term variability, we combine these observations using median statistics which is robust against photometric outliers.
The uncertainty on the combined flux of a single visit is estimated via the weighted average

\begin{equation}
\sigma^{2}_{\rm tot} =  \frac{1}{\sum_{i=1}^k \frac{1}{\sigma^{2}_{i}}}, \label{eqn:err_weight_err}
\end{equation}

where $\sigma_{i}$ are the corresponding uncertainties on the $k$ measured fluxes. This re-sampling makes the long term variability of the light curves more apparent while increasing the signal-to-noise. 
For the purpose of testing ML methods, we only use AGN whose light curves have five or more $\geq5\sigma$ measurements. Although this significantly reduces the sample of AGN, it makes the variability detection and measurement more robust. Furthermore, we focus only on the $\wone$ filter as it is deeper and better time-sampled than $\wtwo$.
Our final sample consists of $8309$ AGN. 
To obtain a training sample, we subsequently classify the light curve of these AGN visually in ``nonvariable'' (7558) and ``variable'' (751). In addition, we split the last category into monotonically increasing (66) and decreasing (98) light curves, while the remaining $587$ vary irregularly (Figure~\ref{fig:wise_lc}).

\begin{figure*}[ht!]
\includegraphics[width=1.02\textwidth]{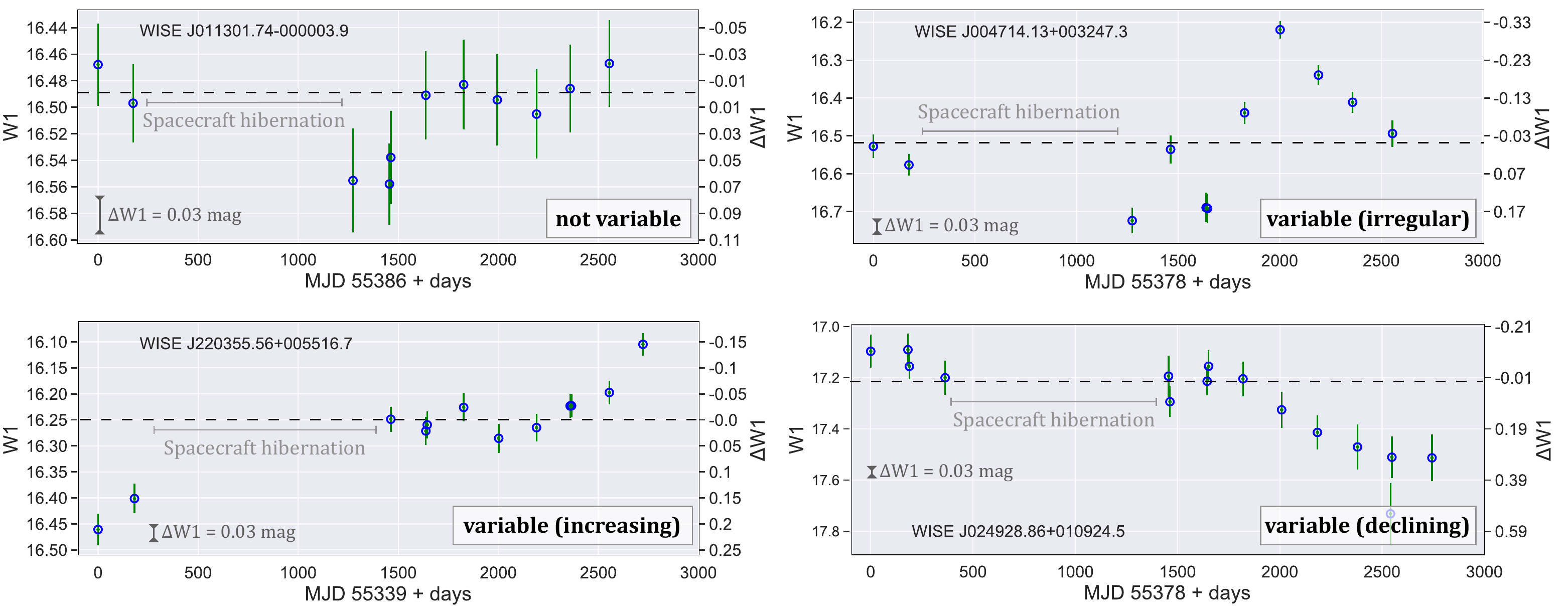}
\caption{Representative examples of \textit{WISE} $\wone$ ($3.4\,{\rm \mu m}$) light curves in the four visual categories. The photometric uncertainties ($\leq$0.03 mag) are indicated for a sense of scale and the dashed line shows the median. Note the different scales of the $y$-axis in the plots.
}
\end{figure*}
\label{fig:wise_lc}

\subsection{Definition of parametric and nonparametric estimators of variability} \label{sec:estimators}

Before applying ML methods, we define a set of estimators to characterize the variability. We distinguish between parametric and nonparametric estimators.

Parametric estimators are derived using the Gaussian Process (GP) framework in Python (\texttt{GPy}\footnote{\url{https://sheffieldml.github.io/GPy/}}). We use a \texttt{RBF} Gaussian kernel and run 100 iterations of optimization on each lightcurve using a \texttt{L-BFGS-B} optimizer, which is typically sufficient for the parameters of the best-fit models to converge. The resulting fit is characterized by a \textit{variance} and a \textit{length scale} parameter. The former is equivalent to a variability estimator, while the latter describes the time scale of a period.
As nonparametric estimators we use the $\chi^2$ test, standard deviation ($\sigma_{w}$), median absolute deviation (MAD), interquartile range (IQR), robust median statistics (RoMS), normalized excess variance ($\sigma^2_{\rm NXS}$), peak-to-peak variability ($\nu$), and the inverse von Neumann ratio ($1/\eta$). These estimators are described in detail in \citet{SOKOLOVSKY17} and we refer to their paper for exact definitions.

All estimators are computed for each of the $8309$ light curves in our sample. The computation of the parametric estimators takes about $30\,{\rm minutes}$ of CPU time on a $3.1\,{\rm GHz}$ processor and the results depend on the random initial conditions for each fit in some single cases ($\sim5\%$). Nonparametric estimators are more advantageous for real-time classification as their computation requires only seconds. Furthermore, their measurement is repeatable.

\section{Finding Variable AGN}\label{sec:classification}

In the following, we identify variable AGN using the unsupervised SOM algorithm implemented in the Python library \texttt{mvpa2}\footnote{\url{http://www.pymvpa.org}} \citep{HANKE09}. Subsequently, we compare its performance to a more commonly used supervised deep-learning multi-layer neural network algorithm that is part of the Python \textit{TensorFlow} package\footnote{\url{http://www.tensorflow.org/}}.

\subsection{Metrics for performance evaluation}

To compare the performance of different ML algorithms as well as the impact of different estimators, we use a common metric known as the confusion matrix, which we here define in its normalized form as

\begin{equation}
\mathcal{C} = \frac{1}{T}
\begin{pmatrix}
{\rm TN} & {\rm FP} \\
{\rm FN} & {\rm TP}
\end{pmatrix},
\end{equation}

where TN, FP, FN, and TP denote true-negative, false-positive, false-negative, and true-positive, respectively, and $T$ is the total sample size (TN$+$TP$+$FN$+$FP).
From this, we derive standard metrics such as purity ($\mathcal{P}$) and completeness ($\mathcal{R}$)\footnote{Note that purity and completeness are equivalent to precision and recall.},

\begin{equation}
{\rm \mathcal{P}} = \frac{{\rm TP}}{ {\rm TP} + {\rm FP} }\,\,{\rm and}\,\,
{\rm \mathcal{R}} = \frac{{\rm TP}}{ {\rm TP} + {\rm FN} },
\end{equation}

as well as the accuracy 

\begin{equation}
{\rm ACC} = {\rm diag}(\mathcal{C}) = \frac{ {\rm TP} + {\rm TN} }{ T},
\end{equation}

the Matthews correlation coefficient \citep[MCC,][]{MATTHEWS75}\footnote{MCC is defined between $-1$ and $+1$ with $-1$ ($+1$) indicating perfect disagreement (agreement) and $0$ meaning the algorithm performs as well as random guessing.}
\begin{equation}
{\rm MCC} = \frac{({\rm TP} \cdot {\rm TN}) - ({\rm FP} \cdot {\rm FN})}{\sqrt{({\rm TP}+{\rm FP})({\rm TP}+{\rm FN})({\rm TN}+{\rm FP})({\rm TN}+{\rm FN})}},
\end{equation}

and the $F_1$ score\footnote{$F_1$ is defined between $0$ and $1$.} defined as 

\begin{equation}
F_1 = 2 \cdot \left( \frac{\mathcal{P} \cdot \mathcal{R} }{ \mathcal{P} + \mathcal{R}} \right).
\end{equation}

The MCC has several advantages over $F_1$ and is generally preferred for assessing the performance of a classification algorithm. For example, MCC does not depend on which outcomes are classified as positive or negative and takes correctly into account TN and FN events.

\begin{figure}
\includegraphics[width=1\columnwidth]{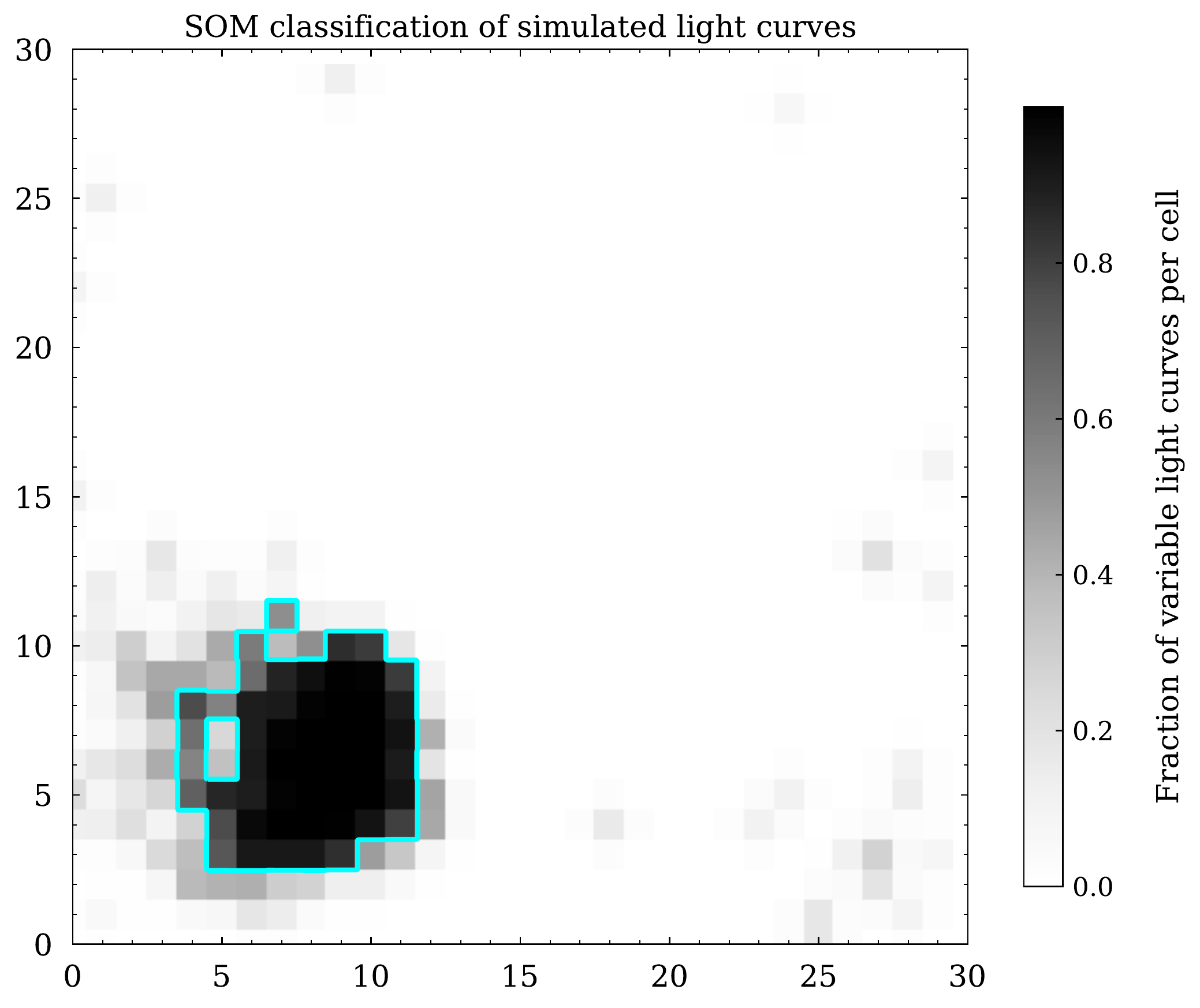}
\caption{Test of our algorithm on simulated light curves. Shown is the fraction of truly variable light curves per SOM-cell (the cyan contour encompasses cells with a fraction higher than $50\%$). We are able to identify variable light curves with a purity of $91\%$ and a completeness of $79\%$.}\label{fig:somsim}
\end{figure}

\subsection{Classification with SOM}
The SOM algorithm reduces an $N$-dimensional dataset (composed of $N$ estimators or parameters) to a two-dimensional grid of $m\times n$ cells. The algorithm preserves topological information as distances in this two-dimensional space map directly to distances in $N$ dimensions. This makes the SOM a powerful tool for visualizing correlations in high-dimensional datasets.
In detail, the SOM algorithm is initialized by the number of iterations ($I$), as well as a length-scale parameter ($\lambda$), learning rate\footnote{The learning rate determines how fast the model is updated per iteration. Commonly, the learning rate is decreased over time for convergence.} ($L_i$), and radius factor ($\sigma_i$). The latter two are decreased with iterations $i$, in the \texttt{mvpa2} implementation of the SOM by the factor $e^{-i/\lambda}$. In the following, we choose as initial values $\sigma_0 = \max(m,n)$ and $L_0 = 0.05$, as well as $\lambda = I/\sigma_0$. For a more detailed review of the algorithm, see, e.g., \citet{MASTERS15}.

\begin{figure*}
\includegraphics[width=1\textwidth]{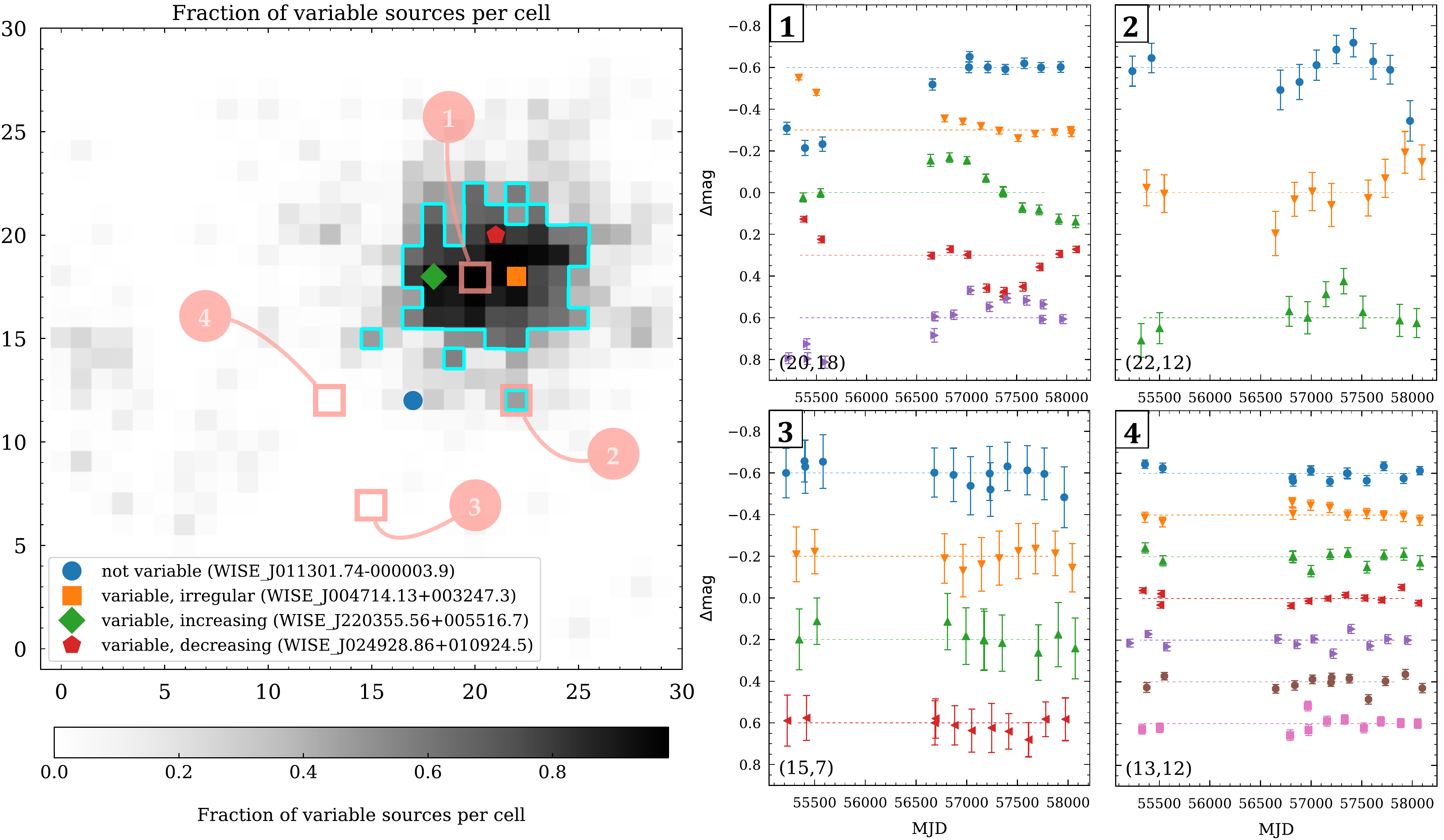}
\caption{Fraction of observed variable AGN light curves per SOM-cell (the cyan contour encompasses cells $>50\%$). The SOM algorithm classifies an AGN as variable with a purity of $86\%$ and completeness of $66\%$. The color symbols indicate the location of the AGN shown in Figure~\ref{fig:wise_lc}. The panels on the right show the light curves (offset by a constant factor) residing in each of the four SOM-cells indicated by the boxes. Cells $1$ and $2$ are dominated by variable AGN, while cells $3$ and $4$ contain mostly nonvariable light curves. All light curves are plotted on the same scale (normalized to median).}  \label{fig:somexample}
\end{figure*}

\subsubsection{Simulations}

We first test the SOM on simulated light curves. For this, we create flat (nonvariable) as well as sinusoidal light curves with varying frequency and phase. These curves are perturbed to achieve similar noise properties as the real photometry and we also apply a time sampling similar to that of real observations. The $7700$ simulated curves include $10\%$ variable light curves, reflecting the visually derived fraction in our flux-limited \textit{WISE} AGN sample.

We calculate all the estimators outlined in Section~\ref{sec:estimators} and normalize and rescale them to their median and a range between $0$ and $1$, respectively.
To train the SOM, we choose a random subsample containing $80\%$ of the total sample (the training sample). We adopt a SOM-size of $30 \times 30$ cells and run $200$ iteration with an initial learning rate of $L_0 = 0.05$. We test different values for the latter two ($50$ iterations and learning rates between $0.005$ and $0.5$) and find changes in the performance of less than $1\%$. The number of cells is chosen to optimize the performance of the algorithm. Specifically, less cells result in a coarser classification, hence a less clear separation of variable and nonvariable light curves. On the other hand, more cells decreases the number of light curves per cell and result in a non-uniform coverage of the map with in a decrease in performance. Overall, we find these choices to be optimal in our case.

Figure~\ref{fig:somsim} shows the fraction of variable light curves in each SOM-cell. The cyan contours encompass cells with a variable fraction of more than $50\%$. The SOM algorithm automatically groups variable light curves around the cells at $(8,7)$ while nonvariable light curves are distributed at larger distances.
We can then quantify the performance of the algorithm by mapping the test sample (the other $20\%$ of the total sample) onto the map. The mapping happens instantaneously for this sample size, which is a strength of the SOM algorithm. Through this mapping, each test light curve gets assigned to a SOM-cell and is then classified as variable if more than $50\%$ of the light curves from the training sample in that cell are variable. This choice of fraction maximizes the metrics and is therefore used throughout this work.
In the following, we assume this binary classification in variable/nonvariable, but note that it is possible to derive a continuous scoring output for each test AGN, which would allow to establish a confidence for each classification.
Using the metrics introduced above, we quantify the success of identifying variable light curves in the test sample as
\begin{equation}
\mathcal{C}_{\rm simulation} =
\begin{pmatrix}
{\rm 0.87} & {\rm 0.01} \\
{\rm 0.03} & {\rm 0.10}
\end{pmatrix},
\end{equation}
with ${\rm ACC}=0.97$, ${\rm MCC}=0.83$, and an $F_1$ score of $0.85$. The purity and completeness of the classification are $91\%$ and $79\%$, respectively, suggesting a ``contamination'' of nonvariable light curves of $9\%$ in a variable sample selected by our algorithm.
Note that the SOMs are randomly initialized, hence these numbers can change for different representations. By running the algorithm multiple times, we find that these changes are on the order of $\pm 0.01$ (or $1\%$ in per-cent notation).

\begin{figure*}[ht!]
\includegraphics[width=1\textwidth]{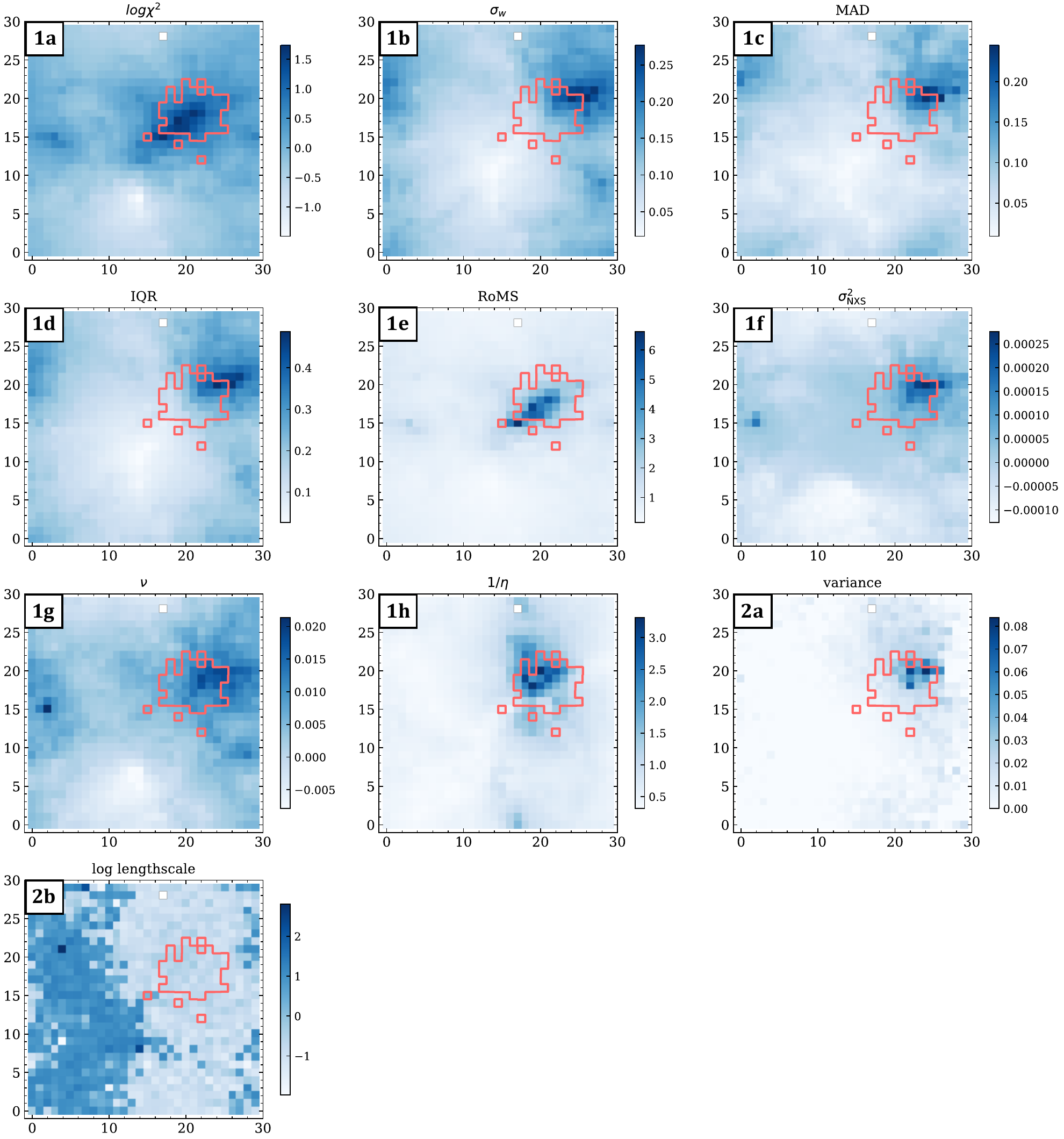}
\caption{Distribution of different estimators on the $30\times30$ cells SOM map. The light-red contours show cells with a variable fraction of $>50\%$. Only nonparametric estimators (panels \textit{1a}$-$\textit{1h}) are used to train the SOM. Most of the estimators correlate well with variability. The estimators $\sigma_{w}$, MAD, IQR, and $\sigma^2_{\rm NXS}$ show offsets indicative of degeneracies in the low signal-to-noise limit.
}  \label{fig:somestmap}
\end{figure*}

\subsubsection{Application to Observed Light Curves}

Having shown that the SOM is a powerful tool to identify variability, we now apply the algorithm to real light curves.
For this, we normalize and rescale the estimators measured for our \textit{WISE} AGN sample as described above. A training fraction of $80\%$ (6647 AGN) is again used to train the SOM. To generate a smooth SOM map, we remove $100$ AGN from the training sample for which at least one estimator lies in the top $1\%$ of the distribution. We find that this clipping improves the performance of the algorithm slightly. Note that this cut is not applied to the test sample. Based on our simulations, we adopt a SOM-size of $30 \times 30$ cells and run $200$ iteration at a learning rate of $0.05$. 

Figure~\ref{fig:somexample} shows the fraction of variable AGN per SOM-cell for the data (the cyan contour encompasses cells with a variable fraction $>50\%$). For educational purposes, we indicate the location of the AGN shown in Figure~\ref{fig:wise_lc} and list on the right light curves contained in the cells at $(20,18)$, $(22,12)$, $(15,7)$, and $(13,12)$. Cells $1$ and $2$ are dominated by variable light curves while cells $3$ and $4$ contain predominantly nonvariable AGN.
In this specific representation of the SOM, variable light curves cluster around the cell at $(21,18)$.
For our basic SOM classification, we find 
\begin{equation}
\mathcal{C}_{\rm SOM} =
\begin{pmatrix}
{\rm 0.85} & {\rm 0.01} \\
{\rm 0.04} & {\rm 0.09}
\end{pmatrix},
\end{equation}
with ${\rm ACC}=0.94$, ${\rm MCC}=0.72$, an $F_1$ score of $0.75$ and a purity (completeness) of $86\%$ ($66\%$).
While the SOM can easily identify variable light curves, we find that splitting into subcategories of variability (i.e., irregular, decreasing, and increasing) can not be achieved robustly. This is not surprising given their small relative number compared to the total sample ($164$ out of $8309$).

By mapping the training sample back onto the SOM cells and computing the median of each estimator per cell, we can visualize and study the correlations of estimators with variability. These elements of the \textit{Kohonen} layer are shown in Figure~\ref{fig:somestmap} (panels \textit{1a}$-$\textit{1h}). The light red contours show cells with a variable fraction $>50\%$ (c.f. cyan contours in Figure~\ref{fig:somexample}). We also show the distribution of the parametric estimators (variance and length scale) on the map (panels \textit{2a} and \textit{2b}).
Most of the estimators correlate well with fraction of variable AGN per cell. Notably, the $\chi^2$, RoMS, and $1/\eta$ estimators correlate best with variability as they peak around the cells with the highest fraction of variable AGN.
Note that the inverse von Neumann ratio ($1/\eta$) is the only estimator discussed here that takes into account the correlation between two successive data points in a time series. Specifically, $1/\eta$ is large for smoothly varying curves, such as smoothly decreasing or increasing light curves. On the other hand, the ratio is small for fluctuations on short time scales (as in highly variable AGN or nonvariable AGN with large photometric uncertainties). 
The other estimators show a wider extent on the maps, suggesting less correlation with variability. This is likely due to degeneracies in the low signal-to-noise regime. Specifically, the estimators MAD, IQR, and $ \sigma^2_{\rm NXS}$ are offset to the north-east and show high values also for nonvariable AGN. We note that the same behavior is seen on the Kohonen maps of the simulated light curves. Such degeneracies may arise because the MAD and IQR estimators do not take into account the photometric uncertainties. As a consequence, a truly nonvariable light curve, poorly sampled in time, can mimic changing brightness (hence a high MAD and IQR) solely due to large photometric errors. Indeed, the average signal-to-noise of the observations is lower in these regions. A similar explanation holds for $\sigma^2_{\rm NXS}$. One could think of removing these estimators for the training of the SOM to improve the identification of variable light curves. We investigate this by training the algorithm only on the $\chi^2$, RoMS, and $1/\eta$ estimators. However, it turns out that overall the performance is slightly worse, suggesting that the removed estimators contain some important information for the classification. Specifically, we find
\begin{equation}
\mathcal{C}_{\rm SOM}|_{\chi^2,\,{\rm RoMS},\,1/\eta} =
\begin{pmatrix}
{\rm 0.86} & {\rm 0.02} \\
{\rm 0.04} & {\rm 0.08}
\end{pmatrix},
\end{equation}
with ${\rm ACC}=0.94$, ${\rm MCC}=0.71$, an $F_1$ score of $0.73$, a purity (completeness) of $84\%$ ($65\%$).

The \textit{variance} parametric estimator shows a good correlation with variability in contrast to the \textit{length scale} estimator, which displays significant scatter and no clear relation. The latter is anti-correlated with the \textit{variance} estimator as expected $-$ it correctly identifies variable AGN with a short length scale (i.e., period), however, the opposite is not true.
Including the \textit{variance} estimator to train the SOM results in a similar performance (${\rm ACC}=0.94$, ${\rm MCC}=0.70$, an $F_1$ score of $0.73$, a purity of $84\%$, and completeness of $64\%$), hence the benefit of including this parametric estimator is questionable, also given that its computation requires two orders of magnitude more CPU time compared to the computation of the nonparametric estimators.
In addition, we test if the performance can be increased by down-sampling the training sample to an equal number of variable and nonvariable AGN (the latter are randomly selected). Indeed, we achieve a higher purity and completeness ($93\%$ and $90\%$) determined on the training sample. However, the performance is worse if determined on the full sample. This is likely because of the small size of the training sample ($1502$ AGN, out of which $751$ are variable). Giving a different weighting to the classes of AGN could improve the performance in future analyses, but implementing this is beyond the scope of this paper.

\subsection{Comparison with deep-learning neural networks}

Finally, we compare the performance of the SOM with a more commonly used supervised deep-learning neural network approach $-$ here the Multilayer Perceptrons (MLP) method implemented in the Python \textit{TensorFlow} package. 
We build a sequential  model with three \textit{Dense} layers. Two of them with 64 and 128 nodes and a rectified linear unit (\texttt{tf.nn.relu}) activation and one with two nodes (yes/no) and normalized exponential (\texttt{tf.nn.softmax}) activation.
The model is compiled using a stochastic gradient descent (SGD) optimizer with a sparse categorical cross entropy.
The deep-learning algorithm is trained on the training sample using the same set of nonparametric estimators as used to train the SOM. We find a confusion matrix of
\begin{equation}
\mathcal{C}_{\rm deep} =
\begin{pmatrix}
{\rm 0.89} & {\rm 0.01} \\
{\rm 0.04} & {\rm 0.05}
\end{pmatrix},
\end{equation}
with ${\rm ACC}=0.94$, ${\rm MCC}=0.65$, and an $F_1$ score of $0.67$. The purity and completeness are $79\%$ and $58\%$, respectively, comparable to the SOM algorithm.
With a similar amount of training time as needed to train the SOM, we find a comparable performance between the two methods. Compared to other optimizers (e.g., the \textit{adams} or \textit{RMSprop} optimizer) we find that the SGD optimizer shows the best performance.
We also test a convolutional neural network (CNN) method and find very similar results.

\section{Conclusions} \label{sec:end}

In this letter, we demonstrate the combination of domain knowledge of how to measure variability with the flexibility and optimization that ML-based approaches bring to large datasets. Using simulated light curves of different variability, we demonstrate how unsupervised self-organizing maps can be used to identify variable AGN light curves in a heterogeneous dataset. This provides powerful means of using ML to identify variability in the presence of photometric noise, selection functions, and heterogeneous sampling in future surveys.
We apply our method to a sample of $8309$ AGN light curves, out of which $\sim10\%$ are identified as variable by our visual inspection. The SOM algorithm can recover variable AGN with a purity of $86\%$ and completeness of $66\%$.
The training of the SOM (done only once) takes less than $100$ seconds ($\sim4000$ objects, $8$ estimators) on a $3.1\,{\rm GHz}$ processor. The classification of a test sample of similar size is instantaneous and can be achieved in real-time for much larger datasets.
In the same CPU time and identical test situations, supervised deep-learning networks perform comparable to the SOM but lack the visualization of the correlation between estimators and the ``fitted'' quantity and cannot easily be applied to datasets with missing labels (i.e., unsupervised).
The SOM framework is powerful to reveal and study connections between variability and other physical properties and processes (e.g., connection between variability and properties of the host galaxy or accretion models).
We here used variable AGN as use-case, but our method can be applied to any light curves to identify supernovae, transiting exoplanets, pulsars, and other transient events in large datasets.

\acknowledgements
\textit{Acknowledgements:} We thank the anonymous referee for the useful comments that improved this paper. We also thank J. Tan for proofreading the manuscript.
This work was supported by Joint Survey Processing (JSP) at IPAC/Caltech which is aimed at combined analysis of Euclid, LSST, and \textit{WFIRST}. This work was supported by Joint Survey Processing (JSP) at IPAC/Caltech 
which was funded by NASA grant 80NM0018F0803.
This publication makes use of data products from the Wide-field Infrared Survey Explorer, which is a joint project of the University of California, Los Angeles, and the Jet Propulsion Laboratory/California Institute of Technology, funded by the National Aeronautics and Space Administration.

\bibliographystyle{apj}
\bibliography{ms}\scriptsize

\end{document}